\documentclass{JHEP3}

\usepackage{graphicx}
\usepackage{amssymb}
\usepackage[mathscr]{eucal}
\usepackage{amsmath}
\usepackage{cite}
\usepackage{afterpage}

\newcommand{\gsimm}{\raise.3ex\hbox{$>$\kern-.75em\lower1ex\hbox{$\sim$}}}
\newcommand{\lsimm}{\raise.3ex\hbox{$<$\kern-.75em\lower1ex\hbox{$\sim$}}}

\newcommand{\be}{\begin{equation}}
\newcommand{\ee}{\end{equation}}
\newcommand{\ba}{\begin{eqnarray}}
\newcommand{\ea}{\end{eqnarray}}

\newcommand{\bea}{\begin{eqnarray*}}
\newcommand{\eea}{\end{eqnarray*}}

\title{Pulsar Constraints on Screened Modified Gravity}

\author{Philippe Brax \\
 Institut de Physique Th\'eorique, CEA, IPhT, CNRS, URA 2306,
 F-91191Gif/Yvette Cedex, France \\ E-mail:
 \email{philippe.brax@cea.fr}}

\author{Anne-Christine Davis\\
 DAMTP, Centre for Mathematical Sciences, University of Cambridge,
 CB3 0WA, UK\\E-mail:
 \email{A.C.Davis@damtp.cam.ac.uk}}
\author{Jeremy Sakstein\\Department of Applied Mathematics and
Theoretical Physics,
Centre for Mathematical Sciences, Cambridge CB3 0WA, UK \\
Perimeter Institute for Theoretical Physics, 31 Caroline St. N,
Waterloo, ON, N2L 6B9, Canada\\ E-mail:
\email{J.A.Sakstein@damtp.cam.ac.uk}}

\abstract{We calculate the rate of energy loss from compact astrophysical
objects due to a scalar field in screened modified gravity models of the
chameleon, dilaton and symmetron types. The cosmological evolution of the field results in a time-variation of the scalar charge
of screened objects implying the emission of scalar radiation. Focusing on binary objects, this leads to an additional decay in
the orbital period complementing that due to the emission of gravitational waves. Using the Hulse-Taylor binary pulsar, the double
pulsar PSR J0737-3039 and the pulsar-white dwarf system PSR J1738+033, we find a new observational bound on the time variation of
the scalar charge of the earth in the Milky Way. We then translate this into a new bound on the range of the scalar interaction in
the Milky Way.  Ultimately, we find that pulsar tests are not competitive with current
observational constraints.}

\begin{document}

\section{Introduction}
Infra-red modifications of gravity \cite{Khoury:2010xi,Boisseau:2000pr} are a
popular explanation for the apparent acceleration of the
Universe\cite{Perlmutter:1998hx, Riess:1998cb}. In many scenarios, this
possibility is realised by augmenting the gravitational sector with a scalar
field coupled to matter. Scalar-tensor theories generically lead to detectable effects in the solar system which have been
constrained drastically by the Cassini probe\cite{Bertotti:2003rm} and other such experiments (see \cite{Will:2001mx} for a
review). Any viable theory that produces interesting cosmological effects therefore requires some sort of \textit{screening
mechanism} that renders the effect of the scalar negligible in dense environments. There are three distinct mechanisms which can
be classified in the Einstein frame using the expansion of the full Lagrangian of the theory up to second order in the small
variation $\delta\phi$ of the scalar field around the background value $\phi_0$
\begin{equation}
\mathcal{L}\supset
-\frac{Z(\phi_0)}{2}(\partial\delta\phi)^2-\frac{m^2(\phi_0)}{2}
\delta\phi^2+\frac{\beta (\phi_0)}{m_{\rm Pl}}\delta\phi \delta T\;.
\label{eq:pertlag}
\end{equation}
The background value $\phi_0$ itself depends on the environment and in
particular on the background matter density. At second order,
the scalar field couples to the variation of the trace of the matter energy
momentum tensor $\delta T$ via the coupling constant $\beta(\phi_0)$.
Both the wavefunction normalisation $Z(\phi_0)$ and the mass of the scalar field $m(\phi_0)$ depend on the environment as does the
matter coupling $\beta(\phi_0)$. All in all, the three effective parameters $Z(\phi_0)$, $m(\phi_0)$ and $\beta(\phi_0)$ are
enough to distinguish the main screening mechanisms.

Here, gravity is modified in as much as the coupling of $\phi$ to matter implies a modification of the geodesics which depend on
the full Newtonian potential $\Phi=\Phi_{\rm N} + \beta\frac{\phi}{m_{\rm Pl}}$ (where $\Phi_{\rm N}$ is the Newtonian potential
satisfying the Poisson equation) in the Einstein frame. The scalar force is screened by the Vainshtein
mechanism\cite{Vainshtein:1972sx}
when $Z(\phi_0)$ is large enough that the matter coupling of the normalised field
$\beta(\phi_0)/Z^{1/2}(\phi_0)$ is negligible\footnote{By which we mean that all appropriate experimental bounds are satisfied.}.
This mechanism is utilised in models of the Galileon type \cite{Nicolis:2008in} and also arises in the decoupling limit of
ghost-free massive gravity\cite{Hinterbichler:2011tt,deRham:2010kj}. The chameleon
mechanism \cite{Khoury:2003aq,Khoury:2003rn,Brax:2004qh} occurs when the mass
$m(\phi_0)$ is large enough to suppress the range of the scalar force in dense
environments. In particular, this implies that the field generated by the bulk
of matter inside a dense body is Yukawa suppressed leaving only the contribution
from a thin shell at the surface of the body reaching the less dense region
outside the compact object. This suppresses the scalar charge of the object relative to its mass thereby rendering the scalar
force negligible compared with the Newtonian one. This
suppression is the essence of the chameleon mechanism. Finally, the symmetron and
dilaton\footnote{This includes the case of the dilaton of string
theory\cite{Damour:1994zq} in the large coupling regime\cite{Brax:2010gi}.}
mechanisms operate by reducing the local value of the matter coupling $\beta(\phi_0)$ to negligible
values\cite{Pietroni:2005pv,Olive:2007aj,Hinterbichler:2010es,Brax:2010gi}. A review of these screening mechanisms can be found in
\cite{Hui:2009kc,Jain:2010ka,Khoury:2010xi,Davis:2011qf,Brax:2012gr,Sakstein:2013pda}. In this work we will be concerned with
models of the chameleon, symmetron and dilaton types only; models of the Galileon type have been
investigated elsewhere \cite{deRham:2012fg,deRham:2012fw}.

The screened models of modified gravity of the chameleon, dilaton and symmetron
types can be described in a unified way\cite{Brax:2012gr} using a tomographic
method\cite{Brax:2011aw} whereby the potential and the coupling function to
matter as functions of the scalar field can be completely reconstructed from the
time dependence of the minimum of the effective potential in the cosmological
background since Big Bang Nucleosynthesis (BBN). This allows one to parametrise
screened models using the cosmological evolution of the scalar mass $m(a)$ and
its coupling strength $\beta (a)$. This description is equivalent to
parametrising modified gravity models using the density dependence of the mass
$m(\rho)$ and the coupling $\beta(\rho)$. A model independent parametrisation
more suited for astrophysical tests can be found in
\cite{Davis:2011qf,Sakstein:2013pda}. Observational constraints on
these theories coming from laboratory and solar system tests
provide bounds on the scalar charge of the earth or equivalently on the
mass of the field in the Milky Way. Astrophysical
constraints \cite{Davis:2011qf,Jain:2012tn,Sakstein:2013pda} bound the mass
and coupling in the cosmological background, which can be related to the
galactic mass of the scalar field using spherical collapse models \cite{Li:2011qda}.

In this work, we will place an independent constraint on the galactic mass of the scalar field
using pulsars in binary systems located in the Milky Way. Pulsars are neutron
stars with Newtonian potentials of $\mathcal{O}(0.01)$. Current constraints
ensure that objects with Newtonian potentials greater than
$\mathcal{O}(10^{-7})$ are self-screening \cite{Jain:2012tn} and so these objects are highly screened. One may then expect no
interesting effects at all, however this is not the case.  Previous works \cite{Silvestri:2011ch,Upadhye:2013nfa} have already
investigated the scalar radiation from a pulsating source where the scalar field is screened. Here we use the fact that
the ambient field value in the galaxy is time-dependent and causes these objects to become more unscreened as time passes. This
gives rise to an increase in the scalar charge. This time-varying charge implies scalar radiation, which causes the period of the
binary system to decrease at a different rate to that predicted by General Relativity (GR). We will find that this gives rise to
new bounds on the range of the scalar force $\lambda_G$ in the Milky Way, thereby placing  direct constraints on the mass
$m_G$ of the scalar field in the local environment.



We calculate the scalar energy change in a finite volume per unit time and
find that it depends on both the gradient of the scalar field and its time
derivative. The gradient is generated by the binary system and leads to a $1/r$
decrease of the field in the local environment. The time dependence is related to the variation of the scalar charge and
eventually the mass of the scalar field
in the galactic halo. This has an effect on the time drift of the orbital period, $P$, of the binary system.
We find that the induced $\dot P$ is small enough to comply with the measured values when the range of
the scalar force in the Milky Way is less than $7.10^5/\beta_G^{1/2} $ Mpc. Models such as \cite{Brax:2012mq,Brax:2013yja} where
both the mass and matter coupling are very large are difficult to probe in astrophysical tests or laboratory experiments
\cite{Mota:2006fz} and so this bound provides a new window into previously unexplored models.
That being said, we ultimately find that tests of the equivalence principle in the
solar system are more constraining that the bounds deduced from binary pulsars and
pulsar-white dwarf systems. 

The paper is arranged as follows. In the next section, we evaluate the rate of
change of the scalar energy in the classical and relativistic
cases focusing on the scalar flux from binary systems in particular.
Only part of this flux has an effect on the gravitational potential energy; most
of it compensates the change of the bulk scalar energy. In the binary case,
this allows us to relate the rate of change of the period $P$ to the local
variation of the scalar charge, which is linked to the range of the
scalar force.
Finally we apply these results to the cases of binary pulsars and pulsar-white
dwarf systems and deduce a bound on the range of the scalar interaction in the
Milky Way. We conclude by discussing this bound in relation to current galactic and solar-system tests of these theories.

\section{Energy flux in Screened Modified Gravity}

\subsection{Classical Scalar Flow}

As a first step, let us analyse the radiation of scalar energy in scalar-tensor
models where the background metric is assumed to be Minkowskian.
We are interested in a scalar-tensor theory where the scalar field is screened
in the presence of dense non-relativistic matter. This theory is characterised
by its bare potential $V(\phi)$ and the coupling function $A(\phi)$
\be
S=\int d^4 x \sqrt{-g}\left(\frac{R}{16\pi G_{\rm N}}
-\frac{(\partial\phi)^2}{2} -V(\phi)\right)+S_{\rm m} (\psi_i, A^2(\phi)
g_{\mu\nu})
\ee
where the particles represented by the fields $\psi_i$ interact with the Weyl
rescaled metric $\tilde g_{\mu\nu}=A^2(\phi) g_{\mu\nu}$.
One of the salient features of the models described here is that
the effective potential of the scalar field in the presence of non-relativistic
matter is modified and becomes
\begin{equation}
V_{\rm eff} (\phi) = V(\phi) -(A(\phi)-1)T
\end{equation}
where $T=-\rho$ is the trace of the conserved energy-momentum
tensor\footnote{The trace of the conserved energy-momentum tensor is related to
the energy-momentum tensor of matter in the Einstein frame via $T_{\rm m}=A(\phi)T$} of
non-relativistic matter. Notice that we have subtracted the contribution from
matter in order to separate the scalar
energy from the matter energy, hence the presence of the $-T$ term.
In this section we concentrate on the case where $\rho$ is time-independent
and localised within a finite volume.
The energy inside this domain is given by
\be
E=\int_D d^3x \left(\frac{ \dot\phi^2 }{2} +\frac{(\nabla\phi)^2}{2} +V_{\rm
eff}(\phi)\right),
\ee
and in this approximation
the scalar field obeys
the Klein Gordon equation
\be
-\ddot \phi +\Delta \phi= \frac{\partial V_{\rm eff}}{\partial\phi}.
\ee
Multiplying by $\phi$ and integrating over the domain $D$ with boundary
$\partial D$, we get
\be
\frac{d}{dt}\int_D d^3 x \left(\frac{\dot\phi^2}{2} + V_{\rm eff}\right)= \int_D d^3 x \dot
\phi \Delta \phi.
\ee
Integrating by parts, we have
\be
\int_D d^3 x \dot \phi \Delta \phi=\int_D d^3 x\left(\partial_i(\dot \phi \partial_i
\phi)-\partial_i \dot \phi\partial_i \phi\right)
\ee
from which we deduce the energy loss due to scalar radiation
\be
\frac{dE}{dt}=\int_{\partial D} d^2S \dot \phi \partial_i\phi n_i
\ee
where we have used Stoke's theorem and $n_i$ is the outward pointing normal to
the boundary ($n^2=1$).
The flux term comes from the current
\be
J_i= \dot\phi \partial_i \phi= T_{0i}.
\ee
In the following, we will generalise this result to relativistic system where
both the matter and gravitational energies must be taken into account. In
particular, we will concentrate
on determining how much of the scalar flux represents a genuine change of the gravitational potential energy for a binary system.
We will show that only a
small part of this flux affects the period.

\subsection{Relativistic Energy Flow}

In this subsection we consider the radiation emitted by a gravitational source in
the Einstein frame of a scalar-tensor theory where the scalar field is locally
screened.
In this frame --- where matter is coupled to the metric $g_{\mu\nu}$ and the scalar field --- the energy emission from a localised
source can be understood
using the
energy momentum pseudo-tensor $t^{\mu\nu}$ defined by Landau and
Lifschitz\cite{Landau:1963zz}, whereby the total conservation law reads
\be
\partial_\mu\left[\ (- g)\ \left( T_{\rm m}^{\mu\nu} + T_\phi^{\mu\nu} +
t^{\mu\nu}\right)\right]=0.
\ee
The matter energy momentum tensor is $T_{\rm m}^{\mu\nu}$ while the one of the
scalar field is $T_\phi^{\mu\nu}$, all evaluated in the Einstein frame. This
conservation law follows directly from the Bianchi identity and therefore the
conservation of the matter and scalar energy-momentum tensors:
\be
D_\mu \left(T_{\rm m}^{\mu\nu} + T_\phi^{\mu\nu}\right)=0
\ee
which shows that energy is exchanged between the scalar and matter
contents of the Universe.
This allows one to identify the total energy density as
\be
T^{00}=(-g)\left( T_{m}^{00} +T_{\phi }^{00} +t^{00}\right)
\ee
and the energy flow vector away from the localised matter system as given by
\be
T_0^i=T_{0\phi }^i+t_0^i
\ee
where we have $g=-1$ as the Universe is locally close to Minkowski space-time
far away from the system.
Integrating over a large volume surrounding the radiating objects, we find that the time-dependence of the total energy is given
by
\be
\frac{dE}{dt}= F_\phi +F_{\rm grav}
\ee
where the total energy inside the volume $D$ surrounding the system is
identified with
\be
E=\int_D d^3 x\ (-g)( T_{m}^{00} +T_{\phi }^{00} +t^{00})
\ee
and the energy fluxes emanating from the objects can be separated into two
components
\be
F_\phi= -\int_{\partial D} d^2S (-g) T_{\phi }^{0i} n_i,\ F_{\rm
grav}=-\int_{\partial D} d^2 S (-g) t^{0i} n_i.
\ee
The gravitational flux $F_{\rm grav}$ corresponds to the gravitational wave
emission while the scalar flux $F_\phi$ results from the dynamics of the scalar
field. Moreover, we have
\be
T_{\phi 0}^i= J^i=\dot \phi \partial^i \phi.
\ee
In the following, we will focus on binary systems in the Milky Way and the flux
across a surface where the gradient $\partial_i \phi$ is given by the one
emanating from
the binary system and not from other local sources. In particular we always
consider that
the Yukawa suppression of the scalar field due to its finite mass $m_G$ in the
local environment of the binary system, e.g. in the Milky Way, occurs on
distances much larger than the distance to the nearest compact object whose
presence disturbs the scalar field generated by the binary system. In practice,
the bound on the range of the scalar interaction in the Milky Way will be large
enough that this is always the case.

\subsection{Scalar Loss of Energy}

Let us evaluate how the scalar energy stored in the domain $D$ varies with time.
For that, let us recall that the energy momentum tensor of the scalar field is
given by
\be
T^{\mu\nu}_\phi= \partial^\mu\phi \partial^\nu \phi
-g^{\mu\nu}\left(\frac{(\partial\phi)^2}{2} +V(\phi)\right)
\ee
and upon using the Bianchi identity we find the Klein-Gordon equation
\be
\Box \phi= \frac{\partial V}{\partial\phi} -\beta \frac{T_{\rm m}}{m_{\rm Pl}}
\ee
where the coupling $\beta$ is defined by
\be
\beta=m_{\rm Pl} \frac{\partial \ln A}{\partial \phi}
\ee
and $T_{\rm m}$ is the trace of the matter energy momentum tensor\footnote{For
non-relativistic matter, the trace $T_{\rm m}=-\rho_{\rm m}$ is related to the
conserved energy density by $\rho_{\rm m}=A(\phi) \rho$.}. We are interested in
the variation of the scalar energy
\be
E_\phi= \int_D d^3x (-g) T^{00}_\phi.
\ee
We focus on gravitational systems where the metric can be put in the
Schwarzschild form
\be
ds^2= -f^{-1} dt^2 + f dx^2
\ee
where, in the Newtonian approximation, $f= 1-2\Phi_{\rm N}$, and $\Phi_{\rm N}$
satisfies the Poisson equation. It will be useful later to note that $\sqrt{-g}= -g^{00}$ for these metrics.
We also neglect the time dependence of $f$, or equivalently of Newton's
potential, and work in this quasi-static approximation where gradients are
larger than time-derivatives of the metric. As a result we find
\be
\frac{dE_\phi}{dt}+\int_{\partial D} d^2S (-g) T_\phi^{0i}n_i= - \int_D d^3x(-g)
\partial^0\phi \beta \frac{T_{\rm m}}{m_{\rm Pl}}+ \int_D d^3 x\partial_i(-g)
\partial^0 \phi \partial^i\phi
\label{cf}
\ee
with no further approximations.
The last term
\be
{\dot E_{\rm flux}}=-\int_D d^3 x\partial_i(-g) \partial^0 \phi \partial^i\phi
\ee
can be equivalently expressed by integrating by parts
to obtain
\be
{\dot E_{\rm flux}}=-\int_{\partial D} d^2 S (-g) T_\phi^{0i}n_i+\int_D d^3x
(-g) \partial^0 \phi \Delta\phi +\int_D d^3x (-g)\partial_i\partial^0\phi
\partial^i\phi,
\ee
which corresponds to the difference between the energy radiated by the scalar
field and the time variation of the bulk scalar energy.
Notice that the flux term cancels now in (\ref{cf}) and we get the energy loss
for the scalar field
\be
\frac{dE_\phi}{dt}=- \int_D d^3x (-g) \partial^0\phi \beta \frac{T_{\rm
m}}{m_{\rm Pl}}-\int_D d^3x(-g) \partial^0 \phi \Delta\phi -\int_D
d^3x(-g)\partial_i\partial^0\phi \partial^i\phi
\ee
where we have defined $\Delta \phi= \frac{1}{\sqrt
-g}\partial_i(\sqrt{-g}\partial^i \phi)$.
The term proportional to $\beta$ corresponds to the exchange of energy with
matter while the last two terms on the right hand side express the intrinsic
variation of energy due to the dynamics of the scalar field.

\subsection{Gravitational Energy Loss}

Using the previous result, we are now in a position to express the loss of
matter and gravitational energy of the system.
We define the total matter and gravitational energy of the system as including
the contributions of both the Landau-Lifschitz and the matter energy momenta
\be
E_{\rm m}=\int_D d^3x (-g)\left( T_{m}^{00}+t^{00}\right).
\ee
The variation of $E_{\rm m}$ is then given by
\be
\frac{dE_{\rm m}}{dt}=-\int_{\partial D}d^2 S (-g) t^{0i} n_i +\int_D d^3x(-g)
\partial^0\phi \beta \frac{T_{\rm m}}{m_{\rm Pl}}+ {\dot E_{\rm flux}}.
\ee
This may be interpreted in the following manner: the matter and gravitational energy
can vary due to the emission of gravitational waves coming from a non-zero $t^{0i}$ and, unlike GR, it
can also change due to the scalar flux. Finally, the matter and gravitational energy can also vary
due to the direct coupling between matter and the scalar field.

In the following we will calculate all these terms in a simple situation where
all the bodies are considered to be point-like and the effects of the scalar
field are screened in each object (note that this is different from the solution for a point-particle, which results in a fully
unscreened field profile).

\section{Time Dependence of the Scalar Field}

The matter and gravitational energy of the system can vary when the scalar field is time-dependent.
There are two origins for the time-dependence of the scalar field. The
first is the motion of the gravitating bodies under study and the
second is the time-dependence of the background scalar field due to the
cosmological or galactic dynamics. In this section, we will analyse the time-dependence of the scalar field due to these two
phenomena.

\subsection{Static Solution}

Let us consider a large domain $D$ where the value of the field is taken to be
the background one $\phi_{\rm G}$ that minimises the effective potential. Screened objects are embedded in this domain and for
the time being, we consider both them and the background field to be static. We also consider that Newton's potential is small
enough
to allow one to approximate the solution outside the objects as Minkowski space.
As a first step, we take a single spherical object in a spherical domain. In
this case, the Klein-Gordon equation reduces to
\be
\frac{d}{dr}\left( r^2 \frac{d\phi}{dr}\right)= r^2 \left(\frac{\partial V}{\partial\phi}
+\frac{\beta \rho}{m_{\rm Pl}}\right).
\ee
This can be written as an integro-differential equation
\be
\frac{d\phi}{dr}= \frac{U(r)}{r^2}
\ee
where the averaged source term is
\be
U(r)=\int_0^r r^2 \left(\frac{\partial V}{\partial\phi} +\frac{\beta
\rho}{m_{\rm Pl}}\right)dr.
\ee
This can be integrated as
\be
\phi(r)= \phi_{\rm G} -\frac{U(r)}{r} -\int_r^\infty \frac{dU}{dr}\frac{dr}{r}
\ee
where we have taken the domain $D$ to be (almost) infinite.
In the following, we will use the scalar charge defined as
\be
Q(r)= 8\pi m_{\rm Pl} \frac{U(r)}{m}
\ee
where $m$ is the mass of the object.

For screened spherical objects, the solution is known to be almost constant
inside the body $\phi=\phi_{\rm c}$ and to follow a Coulombic profile outside.
The solution outside is therefore
\be
\phi(r)= \phi_{\rm G} -(\phi_{\rm G}-\phi_c)\frac{R}{r}
\ee
where $R$ is the size of the object. This corresponds to a scalar charge
\be
Q=\frac{\phi_{\rm G}-\phi_c}{m_{\rm Pl} \Phi}
\ee
where $\Phi= \frac{G_{\rm N} m}{R}$.
When two spherical and screened bodies are embedded in $D$, the approximate
solution is\footnote{The case of an arbitrary number of bodies can treated just
as easily.}
\be
\phi(r)=\phi_{\rm G} - Q_1 m_{\rm Pl} \frac{G_{\rm N} m_1}{\vert r-r_1\vert } -Q_2
m_{\rm Pl} \frac{G_{\rm N} m_2}{\vert r-r_2\vert},
\label{two}
\ee
where
\be
Q_{1,2}= \frac{1}{m_{\rm Pl}(1- \frac{R_1R_2}{r_{12}})\Phi_{1,2}}\left[\phi_{\rm G}
-\phi_{1,2}- \frac{R_{2,1}}{r_{12}}(\phi_G -\phi_{2,1})\right].
\ee
 These have been obtained by imposing that the field values at the surface of the screened objects are $\phi_1$
and $\phi_2$ respectively.
As long as the distance between the bodies is large, $r_{12}\gg R_{1,2}$,  and the values of the field $\phi_{1,2}$ inside the
bodies are small compared to the background value, as befitting the fact that we only focus
 on very dense stellar objects embedded in a galactic environment, we have
\be
Q_{1,2}\approx \frac{\phi_{\rm G}}{m_{\rm Pl} \Phi_{1,2}},
\ee
where $\Phi_{1,2}= \frac{G_{\rm N} m_{1,2}}{R_{1,2}}$ is the Newtonian potential at the surface of each object.
Moreover, we have
\be
\Delta \phi \approx 4\pi m_{\rm Pl} \left[Q_1 G_{\rm N} m_1 \delta^{(3)}(r-r_1)+ Q_2
G_{\rm N} m_2\delta^{(3)}(r-r_2)\right],
\ee
where $\Delta\phi$ is the Laplacian of the Coulombic solution (\ref{two}). This is a good approximation which is also useful when
analysing the time-dependent
case. We deduce that
\be
\frac{1}{r^2} \frac{dU}{dr} \approx 4\pi m_{\rm Pl} \left[Q_1 G_{\rm N} m_1
\delta^{(3)}(r-r_1)+ Q_2 G_{\rm N} m_2\delta^{(3)}(r-r_2)\right],
\label{dd}
\ee
which expresses the fact that $U(r)$ jumps when going across the surface of either
bodies.

\subsection{Time-Dependent Solution}

The time dependence of the scalar solution arises because the bodies sourcing
the scalar field profile do move and also the local background, either
cosmological or galactic, evolves.
In this setting, the Klein-Gordon in the Minkowski approximation for the local metric reads
\be
\Box \phi =\frac{\partial V}{\partial\phi} +\frac{\beta \rho}{m_{\rm Pl}}.
\ee
This is a highly non-linear equation which we can be solved iteratively. The
first step is to use the right hand side evaluated for the static solution of
the previous section, replacing the background value
$\phi_{\rm G}$ by its time-dependent expression $\phi_{\rm G}(t)$ and taking $r_1$ and $r_2$
as time-dependent as well. Because of the screening, we neglect the effect of the
scalar interaction on the
trajectories of the bodies\footnote{This is equivalent to the statement that the scalar charge to mass ratio of each body is
effectively zero. This is justified as the scalar charge for objects with surface Newtonian potentials of order ${\cal O}(0.1)$
must be less than $10^{-15}$, corresponding to a change of Newton's constant $G_{\rm eff}= (1+ \frac{Q_1Q_2}{2}) G_N$ for a binary
system that is less than $10^{-30}$.}. Similarly, the background value $\phi_{\rm G}(t)$ evolves
because the local density of matter $\rho_{\rm G}(t)$ changes with time.
The solution in the domain $D$ can now be expressed in terms of the averaged
potential $U(r,t)$ where the time dependence comes from $\phi_{\rm G}(t)$ and
$r_{1,2}(t)$
\be
\phi(r,t)=-\frac{1}{4\pi} \int_D d^3 r' \frac{1}{r'^2}\frac{ \frac{ \partial U
(r', t- \vert r-r'\vert)}{\partial r'}}{\vert r-r'\vert} -\frac{1}{4\pi}
\int_{\partial D} \frac{\partial}{\partial n} \frac{1}{\vert r-r'\vert}
\phi_{\rm G}\left(r',t-\vert r-r'\vert\right)
\ee
where we have imposed that $\phi=\phi_{\rm G}(t)$ on $\partial D$ and
$\partial/\partial n$ is the derivative in the outward direction.
The boundary integral can be simplified as long as $t\gg R_{\rm G}$ where $R_{\rm G}$ is
the radius of $\partial D$. This is justified since the background field evolves over periods of order the Hubble time and we are
interested in domains
much smaller than the cosmological horizon.
Moreover upon using (\ref{dd}), we find that
\be
\phi(r,t)=\phi_{\rm G}(t) - Q_1\left(t-\vert r-r_1(t)\vert\right) m_{\rm Pl} \frac{G_{\rm N}
m_1}{\vert r-r_1(t)\vert } -Q_2\left(t-\vert r-r_2(t)\vert\right) m_{\rm Pl} \frac{G_{\rm
N} m_2}{\vert r-r_2(t)\vert },
\ee
which can be simplified to
\be
\phi(r,t)=\phi_{\rm G}(t) - \frac{R_1\phi_{\rm G}\left(t-\vert r-r_1(t)\vert\right)}{\vert
r-r_1(t)\vert } -\frac{R_2\phi_{\rm G}\left(t-\vert r-r_2(t)\vert\right)}{\vert r-r_2(t)\vert }.
\ee
We are now in a position to evaluate the energy lost by the system due
to the scalar field.

\section{Energy Loss in the Newtonian Approximation}

\subsection{The Matter and Gravitational Energy}

We will see that the fact that the scalar charge of moving bodies is time-dependent will imply that the matter and gravitational
energy of a system of two
objects varies with time.
In the Einstein frame, the energy momentum tensor of a system of $N$ moving
particles in the non-relativistic approximation is
\be
T^{\mu\nu}_{\rm m} (x^i,t) =\frac{1}{\sqrt{-g}} \sum_{a=1}^N (A_am_a)
\frac{dx^\mu}{d\tau}\frac{dx^\nu}{dt} \delta^{(3)}( x^i- x_a^i(\tau)),
\ee
where the proper time is such that
\be
\frac{d\tau}{dt}=(1+2\Phi_{\rm N} -v^2)^{1/2},
\ee
 $v$ is the velocity of the particles and $\Phi_{\rm N}$ Newton's
potential.
Notice that the mass in the Einstein frame for a particle is given by $m_a A_a$ where
$m_a$ is the conserved mass and $A_a=A(\phi_a)$ is the value of the field inside
the screened particle.
To leading order we find that
\be
(-g) T^{00}_{\rm m}= \sum_{a=1}^N \frac{A m_a v_a^2}{2}\delta^{(3)}( x^i-
x_a^i(\tau))+ \sum_{a=1}^N (A_am_a) \delta^{(3)}( x^i- x_a^i(\tau))
-3 \sum_{a=1}^N (A_am_a) \Phi_{\rm N} \delta^{(3)}( x^i- x_a^i(\tau))
\ee
representing the kinetic energy, the rest mass energy and the potential energy respectively.
We are interested in the matter energy
\be
E_0=\int_D d^3x (-g) T^{00}_{\rm m},
\ee
which contains the potential energy term
\be
\frac{3}{4\pi G_{\rm N}} \int_D d^3 x (\partial \Phi_{\rm N})^2
\ee
where we have used the Poisson equation
\be
\Delta \Phi_{\rm N}= 4\pi G_{\rm N} \sum_{a=1}^N (A_am_a) \delta^{(3)}( x^i-
x_a^i(\tau)).
\ee
The contribution from the Landau-Lifschitz
pseudo-tensor in the Newtonian case is
\be
E_{LL}=\int_D d^3 x (-g) t^{00}= -\frac{7}{8\pi G_{\rm N}} \int_D d^3 x (\partial
\Phi_{\rm N})^2.
\ee
As a result we find that the matter and gravitational energy $E_m=E_0+E_{LL}$ is given by
\be
\int_D d^3x (-g)(T^{00}_{\rm m} + t^{00})= \sum_{a=1}^N ((A_am_a) + \frac{A_am_a
v_a^2}{2}) -\frac{1}{8\pi G_{\rm N}} \int_D d^3 x (\partial \Phi_{\rm N})^2.
\ee
The first term is the rest mass energy, the second one is the kinetic energy and the third one is the gravitational potential
energy.

Let us apply this result to a binary system with semi-major axis $a$.
In this case, the kinetic energy is
\be
\frac{A_1m_1 v_1^2}{2} + \frac{A_2m_2 v_2^2}{2}= \frac{G_{\rm N} (A_1m_1)
(A_2m_2)}{r_{12}}-\frac{G_{\rm N} (A_1m_1) (A_2m_2)}{2a}
\ee
and the gravitational potential energy is\footnote{We have subtracted two
infinite terms coming from the ratio $\frac{\delta(r-r_{\rm i})}{r-r_{\rm i}}$ corresponding to the self-gravitational
contributions to the masses of the point particles.}
\be
-\frac{1}{8\pi G_{\rm N}} \int_D d^3 x (\partial \Phi_{\rm N})^2=-\frac{G_{\rm
N} (A_1m_1) (A_2m_2)}{r_{12}}.
\ee
The total matter and gravitational energy is therefore
\be
E_{\rm m}\equiv \int_D d^3x (-g) (T^{00}_{\rm m} +t^{00})= A_1m_1 + A_2m_2
-\frac{G_{\rm N} (A_1m_1) (A_2m_2)}{2a}.
\ee
The last term is the gravitational energy of the system of two bodies.

\subsection{Energy Exchange}

Let us now investigate the exchange of energy between the scalar field and
matter involving the term
\be
\dot E_{\rm ex}=\int_D d^3x(-g) \beta \frac{T_{\rm m}}{m_{\rm
Pl}}\partial^0 \phi .
\ee
The trace of the energy momentum of matter for a system of particles is
\be
T_{\rm m}= - \frac{1}{\sqrt{-g}} \sum_{a=1}^N (A_am_a) \frac{d\tau}{dt}
\delta^{(3)}( x^i- x_a^i(\tau)).
\ee
As a result, to leading order, we find that the exchange term is given by
\be
\dot E_{\rm ex}= \sum_{a=1}^N m_a \dot A_a - \sum_{a=1}^N \int_D d^3x (m_a \dot
A_a) \left(3\Phi_{\rm N} + \frac{v^2}{2}\right)\delta^{(3)}( x^i- x_a^i(\tau))
\ee
or equivalently after integration
\be
\dot E_{\rm ex}= \sum_{a=1}^N m_a \dot A_a - \sum_{a=1}^N \frac{m_a \dot A_a
v_a^2}{2} -3\sum_{a=1}^N\int_D d^3x (m_a \dot A_a) \delta^{(3)}( x^i-
x_a^i(\tau))\Phi_{\rm N}.
\ee
The first term accounts for the change of the Einstein frame mass, $A_am_a$, when
$A_a$ varies with time. The other terms
\be
\dot E_{A}=\sum_{a=1}^N \frac{m_a \dot A_a v_a^2}{2} +3\sum_{a=1}^N\int_D d^3x (m_a \dot
A_a)\Phi_{\rm N} \delta^{(3)}( x^i- x_a^i(\tau))
\ee
take into account the fact that both the kinetic energy and the potential energy
of the system vary when the coupling function $A$ varies.

\subsection{The Scalar Flux}

We have seen that the scalar field radiates some energy out of $D$ and that some
of it is accounted for by the loss of bulk scalar energy, the difference being
due to
\be
\dot E_{\rm flux}= -\int_D d^3x \partial^0 \phi \partial_i (-g) \partial^i
\phi,
\ee
which is approximately
\be
\dot E_{\rm flux}=-4\int d^3 x \dot \phi \partial_i \Phi_{\rm N} \partial^i
\phi,
\ee
and can readily be calculated in the Newtonian approximation for a binary
system. The integrand is dominated by the behaviour of the fields in the vicinity of $r_1$ and $r_2$. Around both
points we have
\be
\partial_i\Phi_{\rm N} \approx \frac{G_{\rm N} A_{1,2} m_{1,2}}{\vert
r-r_{1,2}\vert^3} (r-r_{1,2})_i
\ee
 and
\be
\partial^i \phi \approx \frac{\phi_{\rm G}(t) R_{1,2}}{\vert r-r_{1,2}\vert^3}
(r-r_{1,2})^i +\frac{\dot\phi_{\rm G}(t) R_{1,2}}{\vert r-r_{1,2}\vert^2}
(r-r_{1,2})^i.
\ee
The time derivative is dominated by
\be
\dot \phi \approx -\frac{\dot \phi_{\rm G} (t) R_{1,2}}{\vert r-r_{1,2}\vert}
\ee
coming from the time dependence of the scalar charge.
The contribution to the change of matter and gravitational energy coming from
the scalar flux reduces to
\be
\dot E_{\rm flux}\approx \frac{ \phi_{\rm G}(t) }{m_{\rm Pl}}\frac{\dot
\phi_{\rm G}(t)}{m_{\rm Pl}} (A_1m_1+ A_2m_2)+ \left(\frac{\dot \phi_{\rm G}(t)}{m_{\rm
Pl}}\right)^2 (A_1R_1 m_1+ A_2R_2m_2),
\ee
which depends on time evolution of the scalar charge only.
The second term gives a dominant contribution when the time derivative term
$R_{1,2} \dot \phi_{\rm G} $ is larger than $\phi_{\rm G}$. For variations of $\phi_{\rm G}$ over
cosmological times, we have $\dot \phi_{\rm G}= {\cal O}(H_0 \phi_{\rm G})$ implying that
the second term is of order ${\cal{O}}(R_{1,2}H_0)\ll 1 $ compared to the first one.  When $\dot \phi_G\gg H_0 \phi_G$ is
larger this may not be the case. We will come back to this point later and we will show in section
5.3 that this is never the case for models of interest.

\subsection{Variation of the Period}

The gravitational potential energy of a binary system is
\be
E_{\rm G}=-\frac{G_{\rm N} (A_1m_1) (A_2m_2)}{2a}.
\ee
Using the third Kepler law stating the constancy of $a^3/P^2$, we find that the
variation of the period $P$ of a binary system can be expressed as
\be
\frac{\dot P}{P}= -\frac{3}{2E_{\rm G}}\left( \frac{dE_{\rm m}}{dt} - m_1 \dot A -m_2 \dot
A+ \frac{G_{\rm N} (A_1\dot A_2+ A_2 \dot A_1) m_1m_2}{2a}\right),
\ee
or, equivalently,
\be
\frac{\dot P}{P}= -\frac{3}{2E_{\rm G}}\left( \frac{G_{\rm N} (A_1\dot A_2+ A_2 \dot
A_1)m_1m_2}{2a}+ \dot E_{\rm flux}- \dot E_A\right).
\ee
The first terms and $\dot E_A$ vanish in the case of screened objects as the
field values $\phi_{1,2}$ are constant inside the objects, depending only on
the density inside the object which is assumed to be constant.
Defining the reduced mass \be\mu= \frac{(A_1m_1)(A_2m_2)}{A_1m_1+A_2m_2}\ee and the
reduced Newtonian potential of the binary system
\be
\Phi_B= \frac{G_{\rm N} \mu}{a},
\ee
the variation of the period of a binary system of screened objects reduces to
\be
\frac{\dot P}{P}= \frac{3}{\Phi_B} \frac{\dot \phi_{\rm G}}{m_{\rm
Pl}}\left ( \frac{\phi_{\rm G}}{m_{\rm Pl}} + \bar R \frac{\dot \phi_{\rm G}}{m_{\rm
Pl}}\right )
\ee
where we have introduced the barycentric radius of the binary system
\be
\bar R= \frac{A_1 m_1 R_1 +A_2 m_2 R_2}{A_1 m_1 + A_2 m_2}.
\ee
We will evaluate this term in the following section.

Let us also comment on the interesting case of a single body moving inside the galaxy. In this case the total matter and
gravitational energy reduces to the kinetic and mass energies
\be
E_m= \frac{Am v^2}{2} + mA
\ee
due to the absence of gravitational binding energy for a single object\footnote{We have subtracted the infinite self gravitational
energy of the particle like in the binary case.}. This is the expected result in the
Newtonian case and in the point-like approximation for the bodies. The balance equation $\dot E_m =\dot E_{ex} + \dot E_{\rm
flux}$ reduces in the approximation that $\dot A=0$ to
\be
\frac{dK}{dt}= Am \frac{\dot\phi_G}{m_{\rm Pl}}\left ( \frac{\phi_G}{m_{\rm Pl} }+ R\frac{\dot\phi_G}{m_{\rm Pl}}\right )
\ee
where the kinetic energy is $K=\frac{Am v^2}{2} $. The particle's speed is affected by the time variation of the environment. In
the case of a static particle, the approximations that we have used break down as one cannot neglect $\dot A$ anymore, which
requires us to go beyond the point-particle approximation.

\section{Phenomenology}

\subsection{Time-variation of the Background Scalar Field}

We have just seen that the energy flux from a binary system due to the scalar
field depends on the time-evolution of the local scalar field
value. In the case of known binary systems, the local environment is the Milky
Way and its galactic halo. An independent test comparing water masers and tip of the red giant branch distances \cite{Jain:2012tn}
places an astrophysical bound that ensures that the Milky Way is self-screening. Hence, the field value $\phi_{\rm G}$ must be
uniformly distributed in the galactic halo
and is related to the scalar charge on earth via
\be
\phi_{\rm G}= Q_\oplus \Phi_\oplus m_{\rm Pl},
\ee
i.e. it is determined by the screening effect in the solar system where
$\Phi_\oplus\approx 10^{-9}$.
The lunar ranging experiment places a bound on a violation of the equivalence
principle in the earth-moon system implying that\cite{Khoury:2003rn}
\be
Q_\oplus\lesssim 10^{-7}
\label{Q}
\ee
for the earth, which gives a tight bound on $\phi_{\rm G}$.

The time variation of the period $P$ of binary systems is then given
by:
\be
\frac{\dot P}{P}\approx 3 \frac{\Phi_\oplus}{\Phi_B} \frac{\dot
\phi_{\rm G}}{m_{\rm Pl}}\left ( Q_{\oplus}+\bar R \dot Q_\oplus\right ),
\ee
which is sensitive to the violation of the equivalence principle parameter
$Q_\oplus$, its time variation and the rate of change of the background scalar field. In the following, we
shall determine a bound on $\dot\phi_{\rm G}/m_{\rm Pl}$ coming from observations of binary systems in the Milky Way.

\subsection{Comparison with Observations}

The time variation of the period $\dot P$ that we have obtained can be applied
to some very precise observations of binary systems, which will allow us to
deduce a  model independent bound on the time variation of the scalar field, $\dot\phi_{\rm G}/m_{\rm Pl}$.
We will consider three cases where $\dot P_{\rm exp}$ has been measured and
compared to $\dot P_{\rm GR}$. The contribution from the scalar field must be
compatible with these measurements and therefore be bounded by the uncertainty
between the experimental value and the GR result. In all cases, the input
parameters are the semi-major axis $a$ and the masses $M_{1,2}$. The masses
$M_{1,2}$ of the two stars can be inferred from the knowledge of two
observables: the rate of advance of periastron $\dot\omega$ and the timing
parameter $\gamma$. It turns out that the calculated values of these observables
is independent of $\dot\phi_{\rm G}$ and deviate from their GR values in a way which
depends on the gravitational charges $Q_{1,2}$ quadratically. For charges
$Q_{1,2}\ll 10^{-7}$ as obtained for neutron stars and white dwarfs, the
deviation from GR is negligible\footnote{Alternatively, one has $A(\phi_{\rm G})\approx 1+\mathcal{O}(Q)$ so that the modified
gravity mass $M_{\rm MG}=A(\phi_{\rm G})M_{\rm GR}$ differs from the GR (Jordan frame) mass by this factor.} and one can therefore
infer the two masses
$M_{1,2}$ and use their values deduced in GR\cite{Damour:1996ke}.

The loosest bound is obtained with the double Pulsar PSR
J0737-3039\cite{Kramer:2006nb} where $a= 878959796.549$ m, $M_1=1.3381(7)
M_\odot$, $R_1\sim 13.72$ km, $M_2=1.2489(7)M_\odot$, $R_2\sim 14.02$ km
$P=0.10225156248(5)$ days, and $\vert \dot P_{\rm exp}\vert =1.252(17) 10^{-12}$
at the 1$\sigma$ level. The GR value is $\vert \dot P_{\rm GR}\vert
=1.24787(13)\times10^{-12}$. The difference is at the $5\times 10^{-15}$ level. As a
result we obtain
\be
\left\vert \frac{\dot \phi_{\rm G}}{m_{\rm Pl}}\right\vert \lesssim 1.5\times 10^{9} H_0.
\ee
A better bound can be obtained using the
Hulse Taylor pulsar PSR B1913+16\cite{Hulse:1974eb} consisting of a pulsar and a
companion neutron star. In this case we have
 $a=1950100$ km, $M_1=1.4398(2) M_\odot$, $R_1 \sim 13.4$ km, $M_2=1.3886(2)M_\odot$, $R_2 \sim 13.5$ km,
$P=7.751939106 $ hours, and $\vert \dot P_{\rm exp}\vert =2.425(1)\times 10^{-12}$ at
the 1$\sigma$ level. The GR value is $\vert \dot P_{\rm GR}\vert
=2.40458(4)\times10^{-12}$. The difference is at the $ 10^{-14}$ level. Hence we
obtain
\be
\left\vert \frac{\dot \phi_{\rm G}}{m_{\rm Pl}}\right\vert\lesssim 6.2 \times 10^{8} H_0.
\ee
The most stringent bound can be obtained with the measurements of the
pulsar-white dwarf binary system PSR J1738+033\cite{Freire:2012mg}.
We have
 $a = 1 738 982 401 $ m, $M_P=1.46^{+0.06}_{-0.05} M_\odot$, $R_P\sim 13.3$ km,
$M_{WD}=0.181^{+0.008}_{-0.007} M_\odot$, $R_{WD}\sim 0.037 R_\odot$,
$P=0.3547907398724(13) $ days, and $\vert \dot P_{\rm exp}\vert =25.9\pm 3.2\times
10^{-15}$ at the 1$\sigma$ level. The GR value is $\vert \dot P_{\rm GR}\vert
=27.7^{+1.5}_{-1.9}\times10^{-15}$. The difference is at the $2\times 10^{-15}$ level. This
implies that
\be
\left\vert \frac{\dot \phi_{\rm G}}{m_{\rm Pl}}\right\vert \lesssim 2 \times 10^{7} H_0.
\ee
This can translated into a bound of the variation of the earth's scalar charge
\be
\vert \dot Q_\oplus\vert \lesssim 2\times 10^{16} H_0.
\ee
 At the surface of the white dwarf this becomes
\be
\vert \dot Q_{WD}\vert \lesssim 10^{12} H_0
\ee
and at the surface of the pulsar
\be
\vert \dot Q_P\vert \lesssim 10^{8} H_0.
\ee

\subsection{Modelling the Time-variation}

The bounds that we have obtained require a study of the dynamics of the scalar
field in the Milky Way. This necessitates a separate study using, for example, a
spherical collapse model such as was presented in \cite{Li:2011qda}. Here we will use a simpler model to determine how
$\phi_{\rm G}$ varies in time; the confirmation of these estimates using the spherical collapse model is left for future work.
Let us assume that the scalar field $\phi_{\rm G}$
in the galactic halo is equal to the value of the minimum of the effective
potential at the galactic halo density.
The time evolution of the local value $\phi_{\rm G}(t)$ for all screened modified
gravity models is then known and is given by\footnote{In the following we will
assume that $A_{\rm G}\approx 1$ in the galaxy.}
\cite{Brax:2011aw}
\be
\frac{d\phi_{\rm G}}{dt}= -\frac{\beta_{\rm G}}{m_{\rm Pl}} \frac{\dot \rho_{\rm G}}{m_{\rm G}^2}
\label{Gn}
\ee
where $m_{\rm G}$ is the mass of the scalar in the local environment and $\beta_{\rm G}$ is
the local coupling.
The time variation of $\phi_{\rm G}$ depends on the variation of $\rho_{\rm G}$ in the
presence of modified gravity. This requires detailed numerical simulations or a
spherical collapse analysis as discussed above. In the following we
will use
a conservative estimate of the time evolution of $\rho_{\rm G}$, i.e. that it varies
over the age of the Universe, hence
\be
\dot \rho_{\rm G} \approx \alpha_{\rm G} H_0 \rho_{\rm G}.
\ee
where we assume that $\alpha_{\rm G}>0$ and of order one.
Using the estimate for $\dot\rho_{\rm G}$ we deduce that for screened models of
modified gravity in the local environment we have
\be
\frac{d\phi_{\rm G}}{dt}\approx -\frac{\alpha_{\rm G} \beta_{\rm G}}{m_{\rm Pl}} \frac{H_0
\rho_{\rm G}}{m_{\rm G}^2}.
\label{dp}
\ee
It is convenient to define $\rho_{\rm G}= 3 H_{\rm G}^2 m_{\rm Pl}^2$ corresponding to the
Hubble rate for $a_{\rm G}=10^{-2}$, i.e. the galactic density is around $10^6$ times
the cosmological dark matter density now.
Hence we find that
\be
\frac{d\phi_{\rm G}}{dt}\approx -3\alpha_{\rm G} \beta_{\rm G} \frac{H_{\rm G}^2}{m_{\rm G}^2} H_0 m_{\rm Pl},
\ee
which we use to evaluate the variation of the period of binary system.
Using the tomographic mapping\cite{Brax:2011aw}, we have the exact
expression
\be
\frac{\phi_G}{m_{\rm Pl}}={9} \int_{0}^{a_G} \frac{\beta (a) \Omega_m (a) H^2(a)}{ a m^2(a)} da
,\ee
which implies that for models where $\beta(a)$ increases with $a$ and $m(a)$ decreases with $a$, as befitting the Damour-Polyakov and chameleon mechanisms respectively, we have the estimate
\be
\frac{\phi_G}{m_{\rm Pl}}= \frac{\beta_G  H^2_G}{\gamma_G m^2_G}.
\ee
The model dependent constant $\gamma_G={\cal O}(0.1)$  can be easily calculated for all known models, for instance for the inverse power law chameleons and  the large curvature $f(R)$ models\cite{Brax:2012gr}.

The variation of the binary period is then given by
\be
\frac{\dot P}{P}\approx 3 \frac{\Phi_\oplus Q_{\oplus}}{\Phi_B} \frac{\dot
\phi_{\rm G}}{m_{\rm Pl}}\left ( 1-3\alpha_G \gamma_G \bar R H_0 \right ).
\ee
 One can check that $\bar R  H_0\ll 10^{-20}$ for all the systems studied here, hence the second term is negligible.
This confirms that the period of binary systems decreases with time. For a single body, this also implies that its kinetic energy decreases due to radiation.

We are now in a position to obtain a bound on the interaction range of the scalar in the galactic environment. First  we have
\be
\frac{\dot P}{P}\approx -9\alpha_{\rm G} \beta_{\rm G} \frac{\Phi_\oplus Q_{\oplus}}{\Phi_B}  \frac{H_{\rm G}^2}{m_{\rm G}^2} H_0 ,
\ee
from which we find
the bound
\be
\left.\frac{m_{\rm G}}{H_0}\right\vert_{\rm pulsar-white \ dwarf} \gtrsim 1.8\times 10^{-3}
(\alpha_{\rm G}\beta_{\rm G})^{1/2}.
\ee
In real space, the range of the fifth force is
\be
\lambda_{\rm pulsar-white \ dwarf} \lesssim 10^6(\alpha_{\rm
G}\beta_{\rm G})^{-1/2}
{\rm Mpc}
\label{puld}
\ee
which corresponds to a very large scale unless $\beta_G$ is large.
\section{Comparison with Other Local Tests of Modified Gravity}

Before concluding, we compare our bounds with other tests of screened modified gravity in both the solar-system and the Milky Way.
Astrophysical bounds (see \cite{Sakstein:2013pda} for more details) constrain the self-screening parameter
\be\chi_0\equiv\frac{\phi_0}{2\beta(\phi_0)m_{\rm pl}},\ee
where subscript zeros refer to cosmological quantities. Cepheid
distance indicators constrain $\chi_0\lsimm 4\times10^{-7}$ \cite{Jain:2012tn}, however here we shall be more conservative for
illustrative purposes and because this bound should be interpreted loosely for models that differ substantially from $f(R)$. By
requiring that Milky Way is self-screening, or equivalently using an independent bound coming from water maser distance estimates
\cite{Jain:2012tn}, we have $\chi_0\lsimm 10^{-6}$ giving a bound on the Compton wavelength of the field in the cosmological
background
\be\lambda_0\lsimm \mathcal{O}(\textrm{Mpc}).\label{eq:lambda_0}\ee
Now since the Milky Way is screened, the field's mass is larger than $m_0$ so that $\lambda_0$ is an upper bound on the range of
the interaction in the galaxy. Comparing (\ref{eq:lambda_0}) with (\ref{puld}) we can see that our new bound is competitive with
these constraints when $\beta_{\rm G}\gtrsim\mathcal{O}(10^{12})$ for $\alpha_{\rm G}\sim\mathcal{O}(1)$.
On the other hand, solar system tests of the equivalence principle (\ref{Q}) imply that
\be
\frac{m_G}{H_0}\gtrsim 10^{10} \left(\frac{\beta_G}{\gamma_G}\right)^{1/2},
\ee
which is stronger even for large $\beta_G$ unless $\gamma_G$ is extremely large. This could be the
case for symmetron models where $\gamma_G \sim (1-a_\star^3/a_G^3)^{-1}$ ($a_\star$ is the scale
factor at the phase transition) is large when $a_G\sim a_\star$ \cite{Brax:2012gr} but this would
require a very unnatural fine-tuning of the symmetron transition time. Therefore, the new bound from
binary systems is not competitive with solar system tests. It is none the less the first bound on
modified gravity deduced from situations where gravity is not weak.

\section{Conclusions}

We have investigated the effects of screened modified gravity on binary pulsar systems and have examined the possible constraints
that can be placed. Pulsars are screened objects and so one would na\"{i}vely assume that their motion is identical to that
predicted in GR. Despite this, we have identified a new and novel effect whereby the time-variation of the field far away from the
system can result in an increase in the scalar charge of the system over time scales of order the Hubble time. This implies that
scalar radiation is emitted by the system and we have calculated the energy-loss due to this effect and have derived its
contribution to the change in the orbital period. Using this, we have been able to place a new and independent bound on the range
of the fifth-force in the galaxy:
\be
\lambda_{\rm G}\lsimm \mathcal{O}(10^6/\beta_{\rm G}^{1/2})\textrm{Mpc},
\ee
where $\beta_{\rm G}$ is the strength of the matter coupling in the galaxy.   This bound is not
as stringent as the one coming from solar system tests. On the other hand, it probes
modified gravity in a very different regime and therefore completes the tests of modified gravity in
situations where gravity is much stronger than in the solar system.


\section{Acknowledgements} We would like to thank P. Freire for correspondence. We are grateful to
the anonymous referees for their various comments and suggestions. JS is grateful to the Perimeter
Institute for Theoretical Physics where part of this work was carried out. Research at Perimeter
Institute is supported by the Government of
Canada through Industry Canada and by the Province of Ontario through the Ministry of Economic
Development \& Innovation. PB
acknowledges partial support from the European Union FP7 ITN
INVISIBLES (Marie Curie Actions, PITN- GA-2011- 289442) and from the Agence Nationale de la Recherche under contract ANR 2010
BLANC 0413 01. ACD is supported in part by STFC.

\section{Appendix}
In this appendix, we will show that there is no scalar field effect on the
binary period at the Newtonian level for unscreened bodies in scalar tensor
theories.
We are interested in scalar field effects in the Newtonian approximation for
unscreened objects with a constant scalar charge and no background evolution.
For a two body system, a good approximation to the solution is
\be
\phi=\phi_{\rm G} - 2\beta m_{\rm Pl} \frac{G_{\rm N} Am_1}{\vert r-r_1(t)\vert}
-2\beta m_{\rm Pl} \frac{G_{\rm N} Am_2}{\vert r-r_2(t)\vert},
\ee
where $A$ and the coupling $\beta$ are time independent. This case corresponds
to the Newtonian approximation for a scalar-tensor theory with no screening
property.
The effect of the scalar field on the period of the system arises from the
integral
\be
\dot E_{\rm flux} =-4 \int_D d^3 x \partial_i\phi \dot \phi \partial^i \Phi_{\rm
N}
\ee
which picks up non-vanishing contributions from the cross terms involving both
particles. Other terms vanish due to rotational invariance.
Such a term is proportional to
\be
m_1 \dot r_1^i \int_D d^3 r (r-r_1)_i \frac{(r-r_1).(r-r_2)}{\vert r- r_1\vert^6
\vert r-r_2\vert^3}.
\ee
The integral is proportional to $(r_1-r_2)^i$ implying that the cross terms are of the form $(m_1 \dot r_1+ m_2 \dot
r_2).(r_1-r_2)$ which vanishes in the centre of mass frame.
Hence we find that the contribution from the scalar field to the variation of the period of a binary system vanishes in the
Newtonian approximation. One must therefore go to the post-Newtonian limit to obtain the
first interesting contribution\cite{Damour:1992we,Damour:1993hw,Damour:1996ke,Damour:1998jk}.

As long as the value of the field in the local environment is considered to be constant, i.e. neglecting the time variation of
$\phi_G$, the scalar radiation in scalar-tensor theories has been extensively
studied\cite{Damour:1992we,Damour:1993hw,Damour:1996ke,Damour:1998jk}. In the case of binary systems, the resulting energy flow,
which follows from the motion of the astrophysical objects, can be either monopolar, dipolar or quadrupolar\cite{Damour:1992we}.
The case of monopolar and quadrupolar radiation is interesting as\cite{Damour:1992we}
\be
\left(\frac{dE_\phi}{dt}\right)^{\rm mono,\ quadru}\propto f_2(Q_A,Q_B) \frac{dE_{\rm grav}}{dt}
\ee
where $\frac{dE_{\rm grav}}{dt}$ is the emission due to gravitational waves and $f_2(Q_A,Q_B)$ is bilinear in the charges of the
two bodies $Q_{A,B}$. As $Q_{A,B}\ll 10^{-7}$ for neutron stars and white dwarfs\footnote{It is interesting to notice that
screened models of modified gravity realise a weakening of the gravitational charge for dense objects $Q\ll \beta_\infty$ which is
due to the existence of a minimum of the effective potential in the presence of matter. When the bare potential is negligible and
the coupling function is of runaway type such as $A(\phi)=\exp(-c\phi^2)$, the field $\phi$ is driven to large values in dense
environments where the gravitational charge becomes large, an anti-screening behaviour which was particularly emphasized in
\cite{Damour:1993hw}.}, the monopolar and quadrupolar fluxes are negligible compared to the
gravitational wave emission.
The most stringent test of scalar emission when the local field has negligible time variation comes from the dipolar emission
rate\cite{Freire:2012mg} for two bodies with different gravitational charges, e.g. for a binary system comprising a neutron star
and a white dwarf\cite{Damour:1992we}
\be
\left(\frac{dE_\phi}{dt}\right)^{\rm dipolar}=-\frac{2\pi}{1+Q_AQ_B}\frac{M_A M_B}{M^2}(Q_A-Q_B)^2\frac{2\pi
G_N(1+Q_AQ_B)M}{P}\frac{1+e^2/2}{(1-e^2)^{5/2}}
\ee
where $M_{A,B}$ are the masses of the objects, $M=M_A+M_B$, $P$ the orbital period and $e$ is the eccentricity.
Observations of white dwarf-neutron star systems such as PSR J1738+033\cite{Freire:2012mg} show that the amount of dipolar radiation is negligible when $Q_{NS}\ll Q_{WD}\ll 10^{-7}$ as
is the case for all neutron stars and white dwarfs. All in all, the monopolar, dipolar and quadrupolar radiation from a screened
system of binary stars is negligible.

For modified theories of gravity where the scalar charge changes with time, i.e $\dot \phi_G\ne 0$, this is no longer the case as
we have seen above.

\bibliography{rad}

\end{document}